\documentclass[10pt,a4paper,twocolumn]{revtex4-1}
\usepackage{amsmath,bm}
\usepackage{graphicx}
\usepackage[latin1]{inputenc}
\usepackage[T1]{fontenc}
\setcitestyle{super}
\usepackage{latexsym}
\usepackage{times}
\newcommand{\dr}{\mathrm{d}r}

\begin{document}

\title{Stimulated Brillouin scattering from surface acoustic waves in sub-wavelength photonic microwires}

\author{Jean-Charles Beugnot\,$^1$, Sylvie Lebrun\,$^2$, Gilles Pauliat\,$^2$, Hervé Maillotte\,$^1$, Vincent Laude\,$^1$, and Thibaut Sylvestre\,$^{1}$}
\affiliation{1: Institut FEMTO-ST, Université de Franche-Comté, CNRS, Besançon, France}
\affiliation{2: Laboratoire Charles Fabry, Institut d'Optique, Universit\'e Paris-Sud, CNRS, Palaiseau, France}

\date{\today}

\begin{abstract}
\noindent Brillouin scattering in optical fibres is a fundamental interaction between light and sound with important implications ranging from optical sensors to slow and fast light. In usual optical fibres, light both excites and feels shear and longitudinal bulk elastic waves, giving rise to forward guided acoustic wave Brillouin scattering and backward stimulated Brillouin scattering. In a subwavelength-diameter optical fibre, the situation changes dramatically, as we here report with the first experimental observation of stimulated Brillouin scattering from surface acoustic waves. These Rayleigh-type hypersound waves travel the wire surface at a specific velocity of 3400 m.s$^{\mathrm{-1}}$ and backscatter the light with a Doppler shift of about 6~GHz. As these acoustic resonances are highly sensitive to surface defects or features, surface acoustic wave Brillouin scattering opens new opportunities for various sensing applications, but also in other domains such as microwave photonics and nonlinear plasmonics.
\end{abstract}

\maketitle

\noindent  
The complex and intriguing dynamics of light and sound interactions in tiny optical waveguides have recently witnessed a renewed interest because of their experimental realization in emerging key areas of modern physics\cite{maldovan_sound_2013}. For instance, the micro and nanostructuring of photonic crystal fibres (PCF) allows for a tight confinement of both photons and phonons, giving rise to new characteristics for Brillouin scattering fundamentally different from those of standard optical fibres. These include the generation of multiple high-frequency hybrid transverse and longitudinal acoustic waves trapped within the small core of microstructured optical fibres\cite{Dainese2006,Kang2009,beugnot_guided_2007,Stiller2011}. Strong photon-phonon coupling has also recently been reported in optical microcavities and new concepts have been introduced such as cavity or surface optomechanics\cite{Kippenberg:Science2008,Matsko2009,Zehnpfennig2011}, Brillouin cooling\cite{Bahl2012a}, on-chip Brillouin scattering\cite{Pant2011}, and microcavity lasers\cite{Grudinin2009,Lee2012,Pant2011}. Moreover, the recent demonstration of simultaneous photonic and phononic bandgaps in nanostructured materials has led to the development of innovative opto-acoustic devices called phoxonic crystals\cite{laude_simultaneous_2011,Laude2005}. In yet another example, tailorable Brillouin scattering was recently reported in nanoscale silicon waveguides\cite{shin_tailorable_2013}. 

Among other microdevices, photonic silica microwires are the tiny and as-yet underutilized cousins of optical fibres \cite{tong_subwavelength-diameter_2003,brambilla_optical_2010}. These hair-like slivers of silica glass, fabricated by tapering optical fibres, enable enhanced nonlinear optical effects and applications not currently possible with comparatively bulky optical fibre. Although microfibres have helped greatly to enhance the optical Kerr effect and stimulated Raman scattering (SRS) for supercontinuum generation\cite{foster_nonlinear_2008,leon-saval_supercontinuum_2004}, stimulated Brillouin scattering (SBS) in these tiny waveguides has not been explored yet. Until now, only transverse acoustic resonances have been reported in long optical fibre tapers with a diameter of a few micrometers\cite{Kang2008a}.

In this work, we present what is to our knowledge the first complete measurement and numerical modeling of Brillouin scattering in photonic silica microwire, revealing the full elastic wave distribution of such long tiny waveguides. In particular, we demonstrate the all-optical generation of a new class of surface acoustic waves (SAWs) and report the first observation of surface acoustic wave Brillouin scattering (SAWBS) in the backward direction. In addition to surface waves, our experimental and theoretical investigations also show that silica photonic microwire exhibits several widely-spaced Brillouin frequencies involving hybrid shear and longitudinal acoustic waves, as previously demonstrated in small-core PCFs\cite{Dainese2006}. Numerical simulations of Brillouin scattering based on the equations of elastodynamics driven by the electrostrictive stress are in excellent agreement with experimental observations.

\begin{figure*}[t]
  \centerline{\includegraphics[width=16cm]{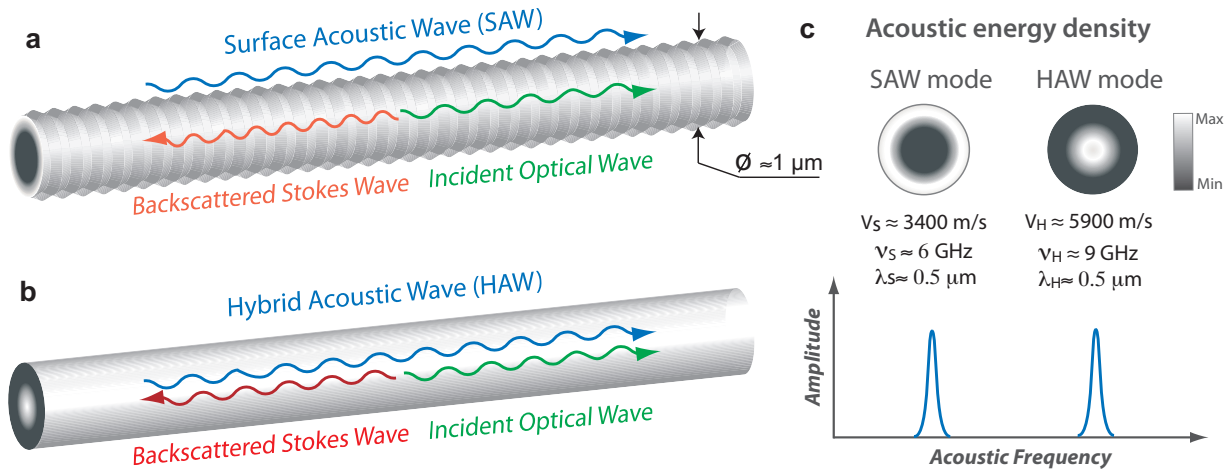}}
  \caption{\textbf{Illustration of surface and hybrid acoustic wave Brillouin scattering in silica microwire:} \textbf{a} Schematic representation of the silica microwire and of the wavevector interaction for surface acoustic waves (SAW). \textbf{b} Same for hybrid (shear and longitudinal) bulk acoustic waves (HAW). \textbf{c} Comparison of kinetic energy densities, acoustic velocities and Brillouin frequency shifts. The SAW kinetic energy is confined below the microwire surface, leading to propagating mechanical ripples, whereas the kinetic energy density of the HAW remains confined within the core without altering the microwire surface.}
\label{fig1}
\end{figure*}

\begin{figure*}[t]
 \centerline{\includegraphics[width=16cm]{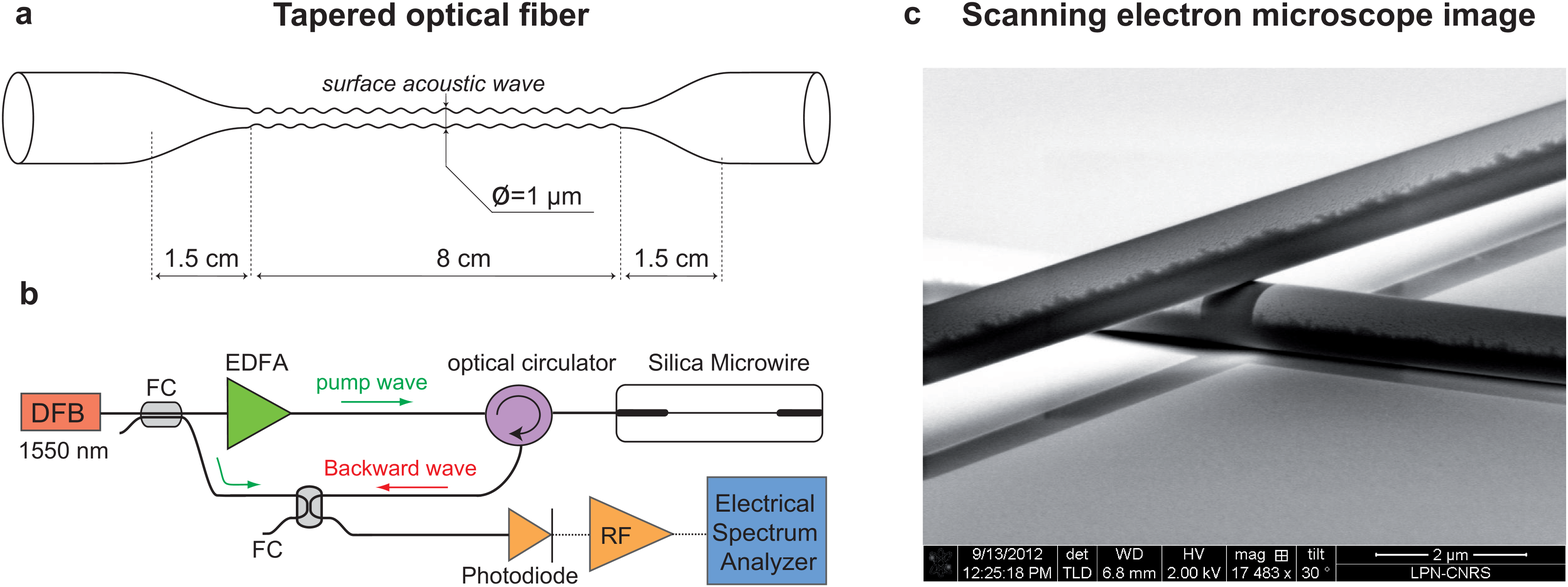}}
  \caption{\textbf{Experiment:}(a) Schematic of the tapered silica microwire. (b) Experimental setup. DFB: Distributed feedback laser. FC: 10:90 Fiber coupler. EDFA: Erbium-doped fibre amplifier. (c) Scanning electron microscope (SEM) image of typical silica wire. Courtesy of  A.-L. Coutrot and C. Dupuis.}
\label{fig11} 
\end{figure*}

When coherent laser light is coupled and guided into a long uniform optical microwire, as shown schematically in Fig.~\ref{fig1}, the light both excites and feels several types of elastic waves with similar wavelengths. In standard optical fibres, light is solely sensitive to shear and longitudinal bulk acoustic waves, leading to well-known nonlinear effects such as guided acoustic wave Brillouin scattering\cite{Shelby1985} (GAWBS) and stimulated Brillouin scattering~\cite{Ippen1972} (SBS), respectively. In contrast, in sub-wavelength optical fibre, the situation changes dramatically, as the guided light and the evanescent field see the outer surface. Light can thus shakes the wire through electrostriction leading to the generation of surface acoustic waves (SAW). The associated propagating surface ripples will then lead to small periodic changes of the effective refractive index along the optical microwire. When passing through this moving refractive index surface grating, light undergoes Bragg scattering in the backward direction according to the phase-matching condition, as in fibre-based SBS from longitudinal acoustic waves (LAW). Backscattered optical wave also experiences a slight shift of its carrier frequency due to the Doppler effect according to photon-phonon energy conservation $\nu={2n_{\mathrm{eff}}V}/{\lambda}$, with $\mathrm{n_{eff}}$ the effective refractive index, $\lambda$ the optical wavelength in vacuum, and $V$ the acoustic phase velocity. The velocity however significantly differs for surface, shear, and longitudinal waves. Surface waves travel at a velocity between 0.87 and 0.95 of a shear wave (for fused silica, $V_S$=3400~m.s$^{\mathrm{-1}}$). This gives rise to new optical sidebands down-shifted from 6~GHz in the light spectrum (Fig.~\ref{fig1}c).  In addition to surface waves, photonic microwires also exhibit hybrid shear and longitudinal acoustic waves due to the tight field confinement, as previously demonstrated in small-core PCFs\cite{Dainese2006}(Fig.~\ref{fig1}b). Hybrid acoustic waves (HAW) propagate at an intermediate speed between shear and longitudinal waves with acoustic frequencies in the range 8-10 GHz. Furthermore, the SAW energy density is confined at the surface of the microwire, leading to mechanical ripples, whereas the HAW energy density remains trapped within the core by the optical force~\cite{BeugnotPRB2012} without altering the wire shape (Fig.~\ref{fig1}c). Both processes are permanently stimulated and amplified by the optical beating between the pump and Stokes waves that mutually reinforces the effects of electrostriction and scattering.

\begin{figure*}[t]
  \centerline{\includegraphics[width=17cm]{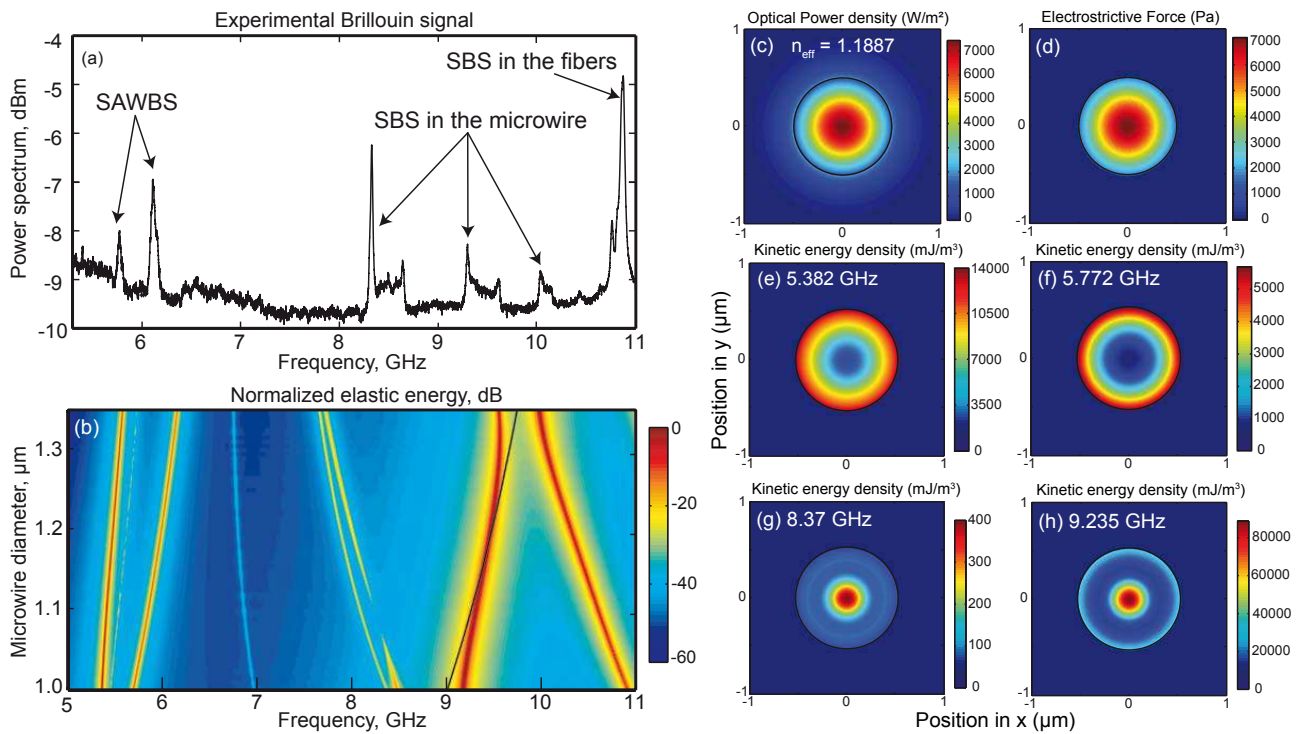}}
  \caption{\textbf{Brillouin scattering in silica microwire:} \textbf{a,} Experimental Brillouin spectrum measured with a c.w.\ laser at 1,550~nm in a 8-cm long silica optical microwire with a diameter of 1$\ \mu$m. See experimental setup for details.  \textbf{b,} Numerical simulation of Brillouin scattering spectrum for a microwire waist varying from 1$\ \mu$m to 1.35$\ \mu$m (See numerical method). The black line shows the phase-matching condition for longitudinal waves ($k_A=2k_p$). \textbf{c-g,} \textbf{c,}  Optical power density in the fundamental TE-like mode at a wavelength of 1,550~nm (n$_{\mathrm{eff}}$= 1.1887). \textbf{d,}  Spatial distribution of the electrostrictive stress under 1~W optical power. \textbf{e, f,} Elastic energy density of resonant surface acoustic waves at 5.382 GHz and 5.772 GHz. \textbf{g, h,} Elastic energy density of hybrid transverse and longitudinal acoustic modes at 8.37 GHz and 9.235 GHz.}
\label{fig2}
\end{figure*}

In our experiment, we investigated a 8-cm long silica optical microwire drawn from a commercial single-mode fibre (SMF) using the heat-brush technique\cite{Shan2013,Baker:11} (for details, see Fabrication method). Fig.~\ref{fig11}c shows an scanning electron microscope image of typical micro/nanowires. The one we used to observe SAWBS is shown schematically in Fig.~\ref{fig11}a. It has a waist diameter of $1\ \mu$m and the input/output tapered fibre sections are 15-mm long. Fig.~\ref{fig11}(b) shows the experimental setup~\cite{Beugnot2008}. As a pump laser, we used a narrow-linewidth continuous-wave distributed-feedback laser running at a wavelength of 1,550~nm. The laser output was split into two beams using a fibre coupler. One beam was amplified and injected in the optical microwire through an optical circulator, while the other beam served for detection. We then implemented a heterodyne detection in which the backscattered light from the microwire is mixed with the input coming from a second coupler. The resulting beat signal was detected using a fast photodiode and averaged Brillouin spectra were recorded using an electrical spectrum analyzer.
Fig.~\ref{fig2}(a) shows the Brillouin spectrum for an input power of 100 mW. Several peaks with different weights and linewidths can be observed in a radio-frequency range from 6 GHz to 11 GHz. First, the high-frequency at 10.86 GHz originates from standard SBS in the 2~m long untapered fibre sections and can be disregarded. More importantly, three other peaks appear at 8.33 GHz, 9.3 GHz, and slightly above 10 GHz, respectively. The two first peaks exhibit a linewidth around 25 MHz, in good agreement with the acoustic phonon lifetime in fused silica ($\sim 10$~ns). Based on numerical simulations described below, we identify them as resulting from Brillouin scattering from hybrid waves while the two other resonances around 6~GHz in Fig.~\ref{fig2}(a) are clearly the signature of surface acoustic waves.

To better understand the light-sound interaction in microwire, we performed numerical simulations based on the equations of elastodynamics extended to account for electrostriction, an effect whereby matter become compressed under the effect of electric field\cite{BeugnotPRB2012,Laude2013} (for detailed explanations, see Numerical method section). With respect to standard three-wave mixing SBS theory\cite{Gaeta1991}, this recent modeling of SBS provides an excellent estimate of the theoretical Brillouin gain spectrum by computing the elastic energy density of acoustic phonons generated by light. In our simulations, we considered the silica microwire as a rod-type cylinder and we took into account all elastic and optical parameters of silica. For the sake of simplicity, we neglected both the tapered and untapered fibre sections. The elastic energy is plotted in Fig.~\ref{fig2}(b) versus acoustic frequency and for a wire diameter varying from $1\ \mu$m to $1.35\ \mu$m. As can be seen, we retrieve most of the surface and hybrid acoustic resonances, as observed in Fig.~\ref{fig2}(a). There are, however, slight differences with experiment regarding the precise resonant frequencies. We attribute them to both the microwire nonuniformity and diameter uncertainty. Nevertheless, we can clearly identify two acoustic modes around 6 GHz, due to SAWBS, and three others around 9 GHz, attributed to SBS. More information is provided by the modal distribution of the interacting waves. Fig.~\ref{fig2}(c) shows the spatial distribution of the optical mode intensity in the waist region at a wavelength of 1,550~nm. The black circle marks out the interface between silica and air. The optical field is guided by the microwire with a rather long evanescent tail extending outside silica. The significant electromagnetic field close to the surface benefits scattering of light by surface acoustic waves. Furthermore, the tight confinement of the optical field into the microwire induces an electrostrictive stress about 30 times larger than in usual optical fibre (up to 7 kPa as compared to 250 Pa under same incident optical power\cite{BeugnotPRB2012}). 
\begin{figure*}[t]
\centerline{\includegraphics[width=17cm]{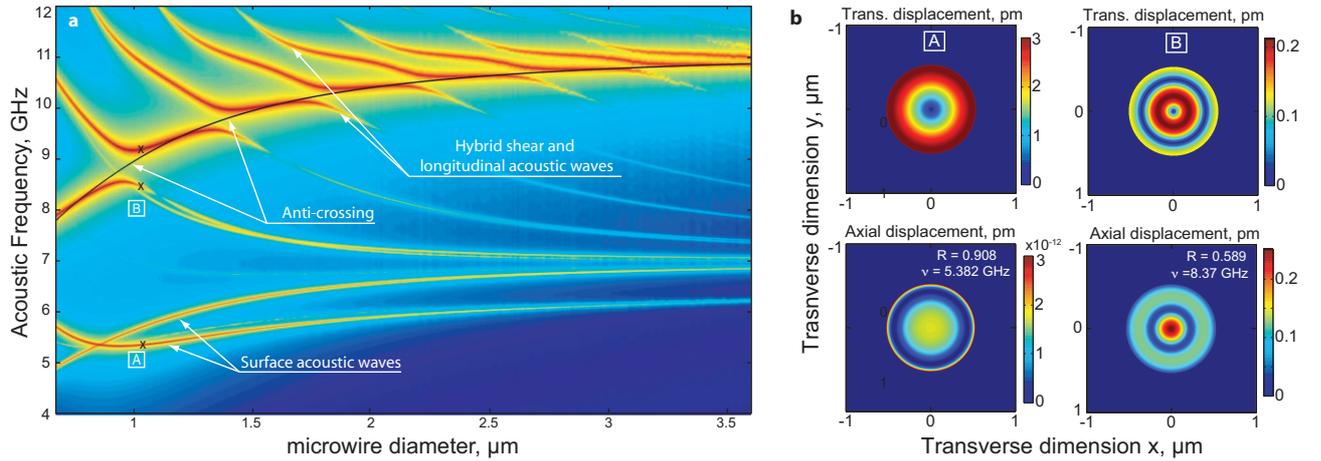}}
  \caption{\textbf{Numerical simulations of full elastic wave spectrum and displacements in silica microwire:} \textbf{a,} Color plot of elastic energy of acoustic waves for a wire diameter varying from 0.6 $\mu$m to 3.5 $\mu$m. The black curve shows the standard phase-matching condition (K=2Kp) for longitudinal waves. White arrows indicate the surface and hybrid acoustic waves, and the anti-crossing points. \textbf{b,} Transverse and axial displacements in the microwire associated with surface and hybrid acoustic modes denoted A and B in \textbf{a}.
  }
\label{fig3} 
\end{figure*}
The electrostrictive stress distribution simply follows the optical mode distribution without extend outward the microwire (Fig.~\ref{fig2}(d)). The kinetic energy density of the two surface acoustic waves at 5.382 GHz and 5.772 GHz are plotted in Figs.~\ref{fig2}(e,f) for a waist of $1.05\ \mu$m. As can be seen, their elastic energy densities are mainly localized below the wire surface. The kinetic energy densities of hybrid acoustic waves at 8.37 GHz and 9.235 GHz, shown in Figs.~\ref{fig2}(g,h), are unlike mostly confined within the core by the optical force. 

To get further insight, we plot in Fig. \ref{fig3}a the full elastic wave spectrum over a wider range of wire diameter till 3.6~$\mu$m. For every diameter, the fundamental optical mode is recomputed and normalized. The refractive index $n_{\mathrm{eff}}$ increases smoothly with the diameter as the optical mode becomes more and more localized within silica rather than air. The two SAWs that we identified before exist for every diameter with a slight shift in frequency. Their Brillouin gain is significant for small core but strongly reduces as the diameter increases beyond the optical wavelength (1.5~$\mu$m). This observation is consistent with the simultaneous decrease of the overlap of SAW with the fundamental optical mode guided inside the core. Fig.~\ref{fig3}a also shows that, for a given microwire diameter, multiple HAW with widely-spaced frequencies can be simultaneously excited, as in small-core photonic crystal fibres~\cite{Dainese2006,Stiller2011}. This means that, in such small sub-wavelength waveguides, the light-sound coupling is very different from standard optical fibres. Brillouin lines are not simply the signature of a single bulk longitudinal phonons. In such tiny waveguides, instead, waveguide boundaries induce a strong coupling of shear and longitudinal displacements, resulting in a much richer dynamics of light interaction with hybrid acoustic phonons. We can also see the appearance of avoided crossings in Fig.~\ref{fig3}a, meaning that HAW on the different branches are not orthogonal and thus can interact. This interaction is strong enough for certain acoustic frequencies to be forbidden at some anti-crossing points. In contrast, the two SAW branches cross without interacting for a diameter $\sim 0.85~\mu$m because they have orthogonal polarizations. To go further into detail, we plotted in Fig.~\ref{fig3}b the associated transverse and axial displacements in the microwire for ones of surface and hybrid modes, denoted A and B in Fig.~\ref{fig3}a. Displacements are simply defined as $\sqrt(u_x^2+u_y^2))$ and ($|u_z|$). R means the ratio between shear and longitudinal displacements (R=1 for a pure shear wave). As can be seen, the surface acoustic mode A at 5.382~GHz exhibits a transversal displacement of a few picometer and a weak axial displacement just below the wire surface. Clearly, this is the signature of a surface Rayleigh wave that combines both a longitudinal and transverse motion to create an elliptic orbit motion. In contrast, the displacements for the hybrid mode B are close (R=0.589) and mainly localized within the wire core.

In conclusion, we reported a clear-cut observation of a new type of Brillouin Stokes radiation induced by surface acoustic waves in sub-wavelength guided optics. This has been achieved in a sub-wavelength diameter optical fiber using the effect of electrostriction. Experimental observations were verified using numerical simulations based on a elasto-dynamics equation. These results show the potential of photonic microwires for surface Brillouin scattering\cite{Comins2001}, optical sensing and nonlinear plasmonics\cite{kauranen_nonlinear_2012,eberle_plasmon_2008}. This work contributes to the further understanding of the intriguing interactions between light and sound in sub-wavelength optics and nanophotonics. Finally, our results constitute the first observation of stimulated Brillouin scattering in tapered fibres with only a few centimeter length of nonlinear propagation.
\section*{Methods}

\footnotesize

\subparagraph*{\hskip-10pt Fabrication method}

The silica microwire has been tapered from a standard telecom fibre (SMF-28) using the heat-brush technique\cite{Baker:11}. The fibre to be pulled is attached at two computer-controlled translation stages. The fibre is softened on its central part with a small butane flame. The gas flow is monitored and regulated by a mass-flow controller. The flame is kept fixed while the two translation stages elongate the fibre to create the microwire. The microwire shape is fully controlled by the trajectories of the two translation stages. The un-tapered parts, upstream and downstream the microwire, allow a very easy light injection and collection. During the pulling, a laser light is coupled inside the microwire and collected at the output end to control several parameters (transmitted power, mode shape, and spectrum). With this home-made pulling plateform we routinely achieved light transmission larger than 90\% over the full fibre including the tapered and un-tapered parts, even with microwire diameters of a few hundred nanometers and lengths up to several centimeters.

\subparagraph*{\hskip-10pt Numerical method} 
Our experimental observations of surface and hybrid acoustic waves are modelled with the equations of elastodynamics including the electrostrictive stress induced by the optical field. We must emphasize that radiation pressure is negligible in our case.  Specifically, the displacements $u_i$ in silica microwire are given by a simple partial differential equation which reads:
\begin{equation}
\rho \frac{\partial^2 u_i}{\partial t^2} - [c_{ijkl} u_{k,l}]_{,j}  = - T_{ij,j}^{es} ,
\label{eq3}
\end{equation}
where $c_{ijkl}$ is the rank-4 tensor of elastic constants. $T_{ij}^{es} =  -\epsilon_0 \chi_{klij} E_{k} E^{*}_{l}$ is the eletrostrictive stress tensor, with the rank-4 susceptibility tensor $\chi_{klij}=\varepsilon_{km} \varepsilon_{ln}  p_{mnij}$ and $p_{mnij}$ the elasto-optic tensor. $\epsilon_0$ is the permitivity of vacuum. The force term with detuning frequency $\omega = \omega_1 - \omega_2$ is proportional to $E^{(1)}_{k} E^{(2)*}_{l}\exp(i(\omega t - k z))$ with $k = k_1 - k_2$. Here we assume that the total optical field results from the superposition of the pump and Brillouin Stokes waves with angular frequencies $\omega_{1,2}$ and axial wavevectors $k_{1,2}$. If the two optical waves are propagating in opposite directions (See Figs.~\ref{fig1}a,b) $k \approx 2 k_1$ and we speak of backward SBS or SAWBS. We also consider in our model the phonon lifetime by taking into account the elastic losses assuming a complex tensor. This loss model is compatible with the usual assumption that the product of the quality factor Q and the acoustic frequency is a constant for a given material (e.g., for silica, Q$\times$f =5 THz). Further applying Green's theorem to Eq. 1, we get
\begin{equation}
-\omega^2 \int_S \rho v^*_i u_i + \int_S v^*_{i,j} c_{ijkl} u_{k,l} = \int_S \dr v^*_{i,j} T^{\textrm{\scriptsize{es}}}_{ij},
\end{equation}
which accounts for the theorem of virtual work for the elestrostrictive stress. For numerical computations, we used the Galerkin nodal finite element method (FEM) to transform the integral equation into the following linear matrix system
\begin{equation}
(K(k) - \omega^2 M) U = X(k) T^{\textrm{\scriptsize es}} ,
\label{eq5}
\end{equation}
with mass matrix $M$, stiffness matrix $K(k) = K_0 + k K_1 + k^2 K_2$, and $X(k)=X_0 + k X_1$.
$U$ is the vector of nodal displacements $\bar{u}_i$.
Solutions of equation~(\ref{eq5}) as a function of frequency detuning yield rigorous distribution of displacements within microwire cross-section as shown in Fig.~\ref{fig2}. The associated energy density can be directly compared to experimental measurements of Brillouin scattering. Finally, the electrostriction stress tensor is defined by the optical modal distribution (Figs.~\ref{fig2}c,d), which is beforehand calculated using a finite element method (Comsol).

\bibliographystyle{nature2}
\bibliography{Brillouin-2}

\bigskip

\section*{Acknowledgements}
\noindent This work was supported by the ANR LABEX Action and the Fonds europ\'een de d\'eveloppement \'economique r\'egional  (FEDER) under contract INTERREG-IVA program. J.C.B. and T.S. acknowledges the support of the University of Franche-Comt\'e and the BQR PRES. The authors thank A.-L. Coutrot and C. Dupuis from LPN in Marcoussis for providing the wires image.
\section*{Author Contributions}
\noindent J.C.B. performed the main experimental observations and characterizations. T.S. wrote the main manuscript text and supervised the overall project. S.L. and G.P. designed and fabricated the optical microwires. V.L. and J.C.B. made the numerical modeling. H.M. launched the initial scientific project. All authors discussed the results and substantially contributed to the manuscript.

\section*{Additional information}
\noindent The authors declare no competing financial interests. Correspondence and requests for materials should be
addressed to to T.S. or J.C.B. (email: thibaut.sylvestre@univ-fcomte.fr and jean-charles.beugnot@femto-st.fr).
\end{document}